%% file: booii.tex
\documentclass[iop]{emulateapj}
\newcommand{\Booii}{Boo~II}

\newcommand{\Bootesii}{Bo\"otes~II}
\newcommand{\BOOTESII}{BO\"OTES~II}

\newcommand{\StarA}{SDSS\,J135751.2+125137.0} 
\newcommand{\StarB}{SDSS\,J135759.7+125426.4} 
\newcommand{\StarC}{SDSS\,J135756.2+125207.7} 
\newcommand{\StarD}{SDSS\,J135801.4+125105.2} 

\newcommand{\Gehaprep}{Geha et al. (in prep)}
\newcommand{\kms}{km\,s$^{-1}$}
\newcommand{\logg}{$\log g$}
\newcommand{\micro}{$\nu_{\rm micr}$}

\usepackage{hyperref}
\usepackage{amsmath}
\usepackage{amssymb}
\usepackage{graphicx}

\begin{document}
\shorttitle{High-Resolution Spectroscopy of \Booii}
\title{High-Resolution Spectroscopy of Extremely Metal-Poor Stars in
  the Least Evolved Galaxies: \BOOTESII\altaffilmark{*}}

\author{Alexander P. Ji\altaffilmark{1}, Anna Frebel\altaffilmark{1,2}, 
  Joshua D. Simon\altaffilmark{3}, Marla Geha\altaffilmark{4}}

\altaffiltext{*}{This paper includes data gathered with the 6.5 m Magellan Telescopes
located at Las Campanas Observatory, Chile.}
\altaffiltext{1}{
  Department of Physics and Kavli Institute for Astrophysics and Space
  Research, Massachusetts Institute of Technology, Cambridge, MA
  02139, USA; \texttt{alexji@mit.edu}}
\altaffiltext{2}{
  Joint Institute for Nuclear Astrophysics - Center for Evolution of the Elements, East Lansing, MI 48824}
\altaffiltext{3}{Observatories of the Carnegie Institution 
                 of Washington, 813 Santa Barbara St., 
		 Pasadena, CA 91101}
\altaffiltext{4}{Astronomy Department, Yale University, New Haven, CT 06520}

\begin{abstract}
We present high-resolution Magellan/MIKE spectra of the four brightest
confirmed red giant stars in the ultra-faint dwarf galaxy \Bootesii\ (\Booii).
These stars all inhabit the metal-poor tail of the \Booii\ metallicity
distribution function.
The chemical abundance pattern of all detectable elements in these stars
is consistent with that of the Galactic halo. However, all four stars
have undetectable amounts of neutron-capture elements Sr and Ba, with
upper limits comparable to the lowest ever detected in the halo or in
other dwarf galaxies.
One star exhibits significant radial velocity variations
over time, suggesting it to be in a binary system. Its variable
velocity has likely increased past determinations of the \Booii\
velocity dispersion.
Our four stars span a limited metallicity range, but their enhanced
$\alpha$-abundances and low neutron-capture abundances are consistent
with the interpretation that \Booii\ has been enriched by very few
generations of stars.
The chemical abundance pattern in \Booii\ confirms the emerging trend
that the faintest dwarf galaxies have neutron-capture abundances
distinct from the halo, suggesting the dominant source of
neutron-capture elements in halo stars may be different than in
ultra-faint dwarfs. 
\end{abstract}

\keywords{galaxies: dwarf --- galaxies: individual (\Booii) --- Local
  Group --- stars: abundances}

\section{Introduction}\label{s:intro}
The ultra-faint dwarf galaxies (UFDs) in orbit around the Milky Way
are important probes of several extreme regimes.
They are the smallest and the most metal-poor galaxies, falling at the
bottom of the luminosity-metallicity relation \citep{Kirby08,Kirby13b}.
As the most dark-matter dominated systems known
\citep{Simon07,Strigari08,Simon11}, UFDs are 
promising targets in the search for a dark matter annihilation
signal (e.g., \citealt{DrWag15}).
There are hints that the stellar initial mass function in
dwarf galaxies is bottom-light, suggesting differences in the nature of
early metal-poor star formation \citep{Geha13}.
The UFDs are distinctly tied to reionization, both
as important sources of ionizing photons \citep{Wise14,Weisz14c} and
as victims of ionizing radiation \citep{Brown14}.
Stars stripped from dwarf galaxies may be important contributors to the
metal-poor tail of the galactic stellar halo \citep{Kirby08,Frebel10a,Frebel15}.
Furthermore, the most metal-poor of these UFDs may be relics of the
era of first galaxies, taking a snapshot of the first stages of chemical
enrichment and potentially opening up a window to study the first stars
and galaxies \citep{Frebel12,Frebel14,Ji15}.

\Bootesii\ (hereafter \Booii) is a UFD discovered in the Sloan Digital
Sky Survey \citep{Walsh07}.
At a distance of 42 kpc, it is one of the closest UFDs known \citep{Walsh08}. Its
size, luminosity, and average metallicity show it to be at the very
low end of the size-luminosity and luminosity-metallicity relations (\Gehaprep),
comparable to systems like Segue~1, Willman~1, Ursa~Major~II, and
Coma~Berenices (e.g., \citealt{Martin08,Kirby13b}).
Photometric and medium-resolution spectroscopic observations have
further established that \Booii\ has an extreme
mass-to-light ratio consistent with other UFDs
(\citealt{Walsh08,Martin08,Koch09}, \Gehaprep).
However, several important questions require chemical abundances from
high-resolution spectroscopy, ideally for multiple stars in the
system. In particular, as \Booii\ is one of the smallest galaxies, it
is a candidate to be a relic of one of the first galaxies, like
Segue~1 \citep{Frebel14}.

Here, we present chemical abundance results from high-resolution
spectroscopy of the four brightest stars in \Booii.
In Section~\ref{s:obs}, we describe our observations.
Section~\ref{s:abund} details our abundance analysis
procedure and compares the chemical signature of \Booii\ to that of
other UFDs and to metal-poor stars in the halo.
One of the stars in \Booii\ appears to be a binary system, which we
investigate in Section~\ref{s:rv}.
We summarize our results and conclude in Section~\ref{s:concl}.

\section{Observations}\label{s:obs}
\begin{figure}
  \includegraphics[width=8cm]{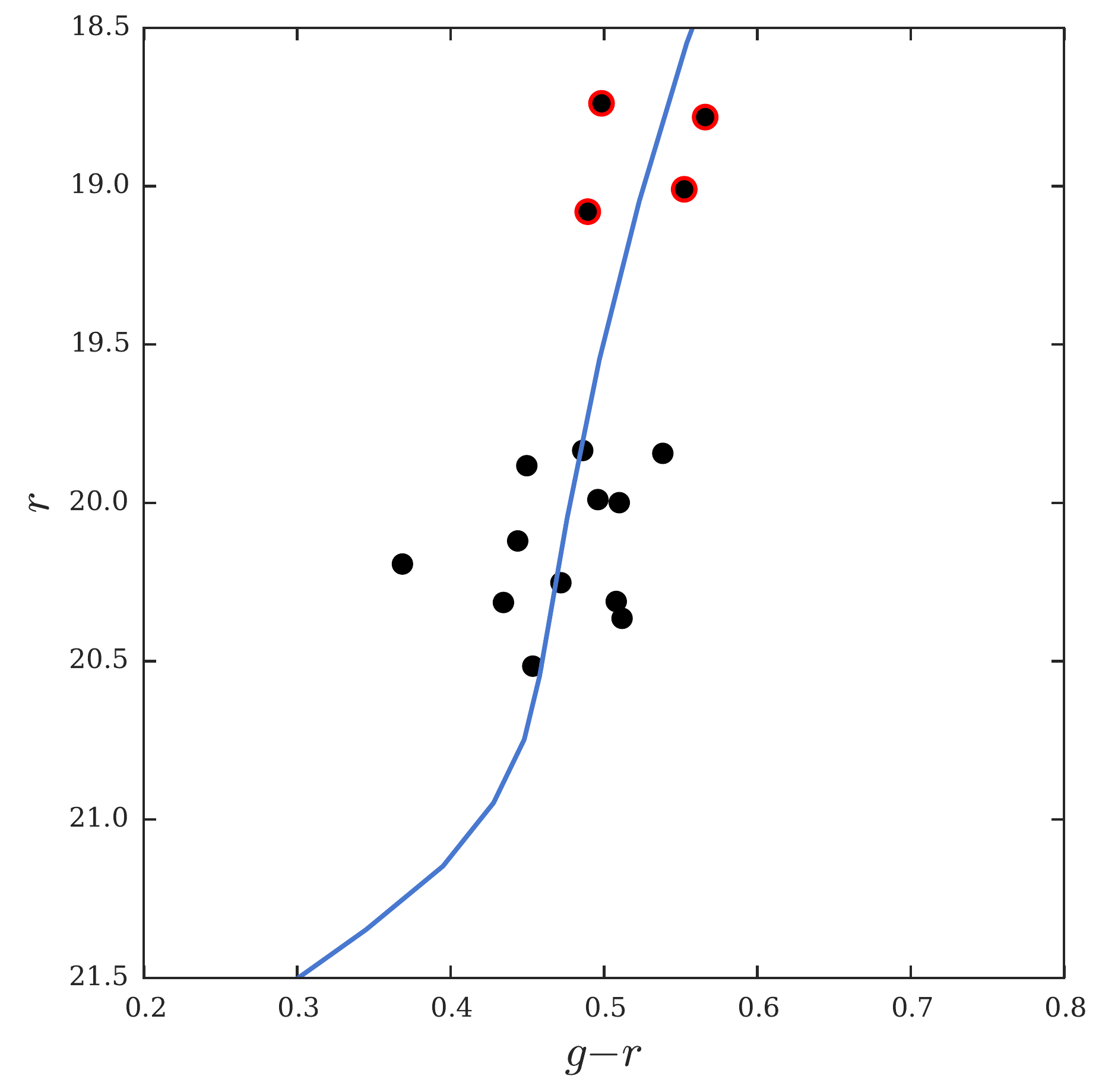}
  \caption{\Booii\ color-magnitude diagram with SDSS DR12 photometry
    of member stars (\Gehaprep, \citealt{SDSSDR12}).
    Red circles indicate the four stars observed with MIKE.
    Blue line is the M92 ridgeline \citep{Clem08} shifted to the
    \Booii\ distance of 42\,kpc.
    Followup photometry has confirmed that these four stars lie
    closer to the M92 ridgeline than is apparent from the SDSS
    photometry (\Gehaprep).
  \label{f:cmd}}
\end{figure}

A total of 16 \Booii\ member stars were identified with Keck/DEIMOS observations
(\Gehaprep).
We selected the four brightest members on the red giant branch from
the \Gehaprep\ sample for high-resolution followup (see
Figure~\ref{f:cmd}). Despite being 
the brightest members, all four stars push the limit of
high-resolution spectroscopy with current telescopes. Two of the 
stars have $V \sim 18.9$ and two have $V \sim 19.2$.
The next-brightest member stars are over 1 magnitude fainter.
We label the stars by their Sloan Digital Sky Survey coordinates.

\begin{figure*}
  \begin{center}
  \includegraphics[width=7cm]{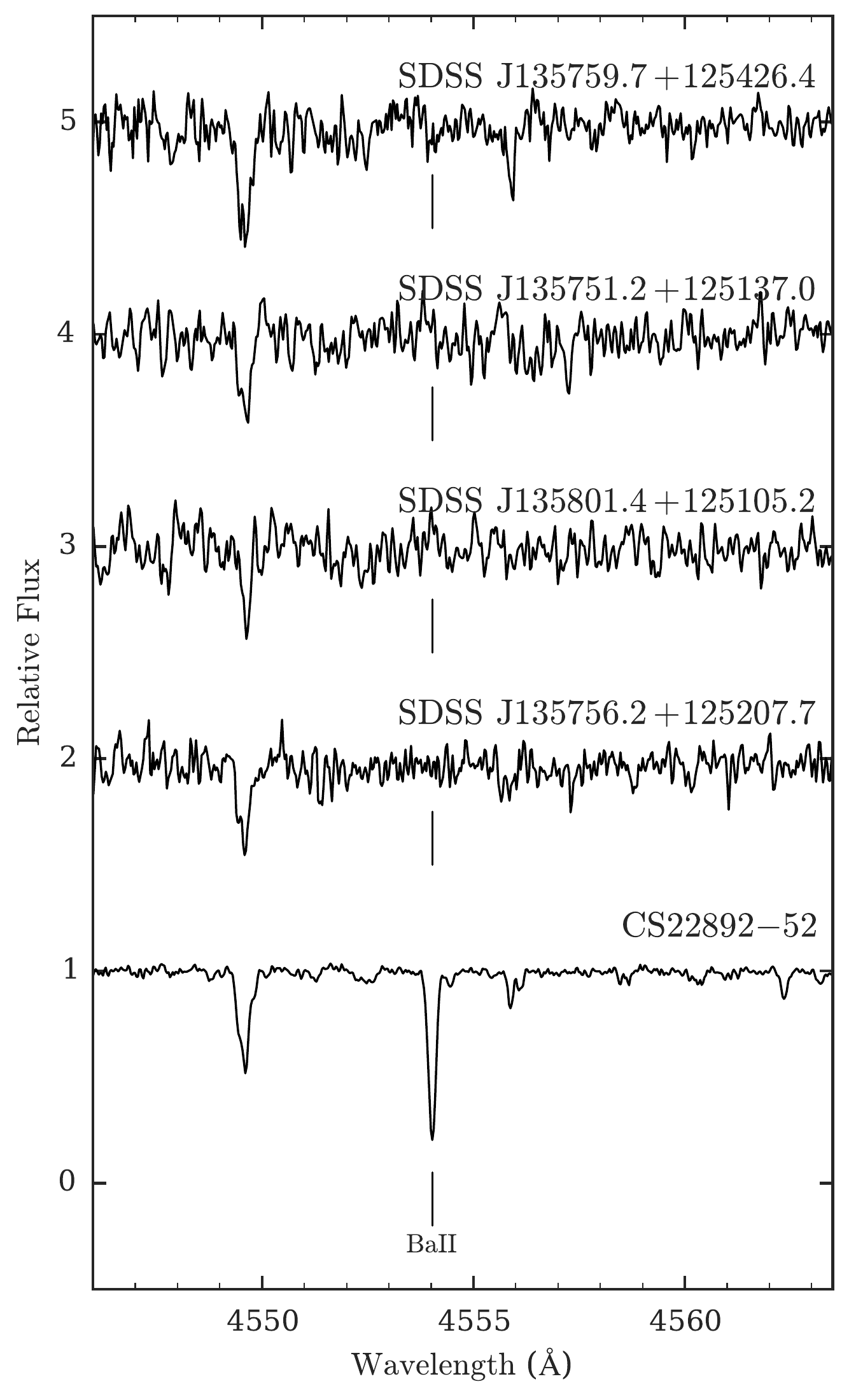}
  \includegraphics[width=8.28cm]{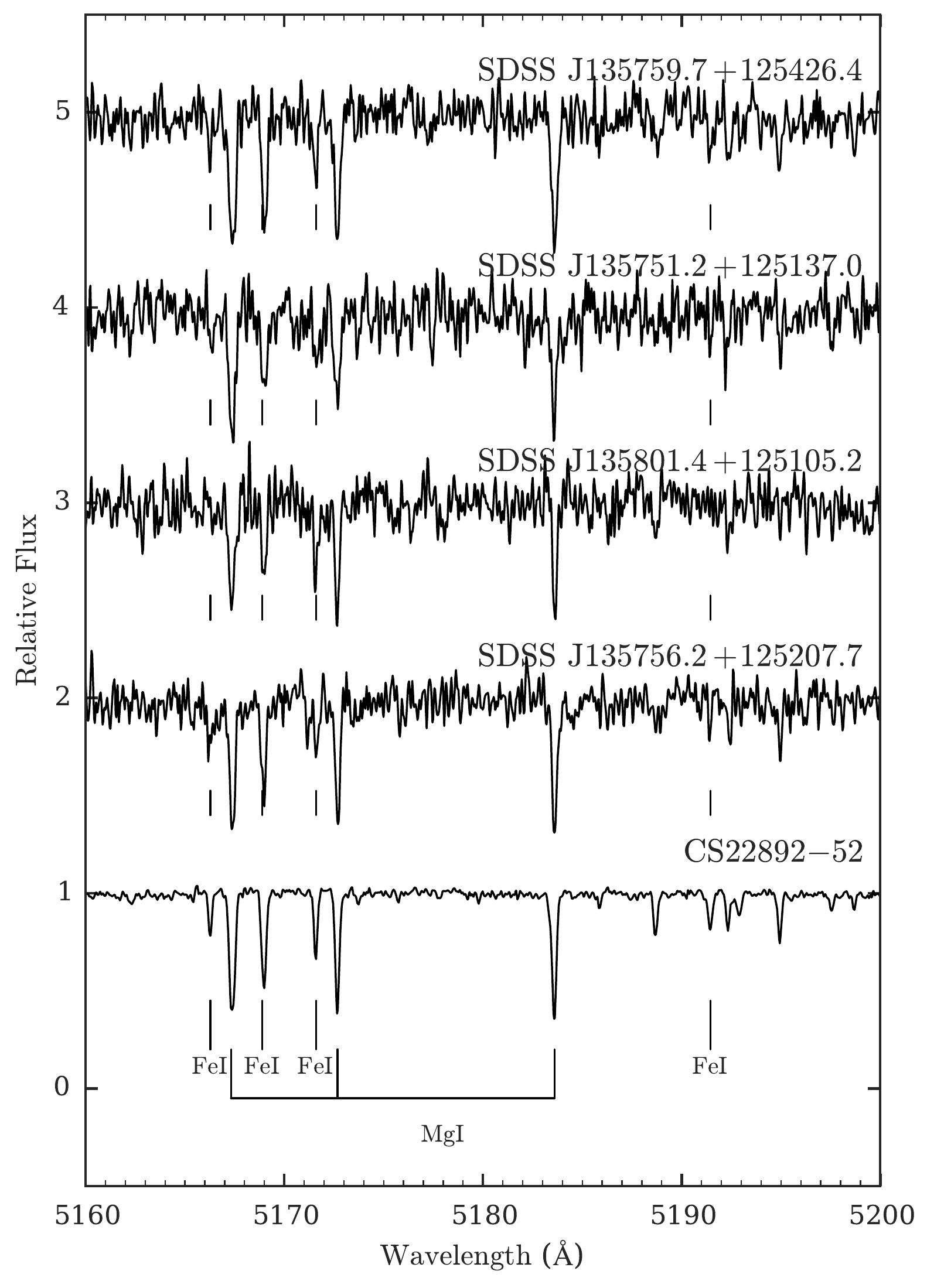}
  \end{center}
  \caption{Magellan/MIKE spectra for our four \Booii\ stars, shown
    near the 4554\,\AA\ Ba\,II line and the Mg b triplet lines around 5180\,\AA.
    For comparison, we show the r-process star CS22892$-$52.
    The four \Booii\ spectra are arranged in order of $\mbox{[Fe/H]}$.
    No Ba\,II is detected in any of the \Booii\ stars.
    \label{f:specoverview}}
\end{figure*}

The four target stars were observed with the Magellan Inamori Kyocera
Echelle (MIKE) spectrograph \citep{Bernstein03} on the Clay telescope. 
MIKE covers the full optical wavelength range, from 3500\,\AA\ to
9000\,\AA.
Details of the observations are shown in Table~\ref{tbl:obs}, and
important spectral regions are shown in Figure~\ref{f:specoverview}.
Three of our stars were observed with a 1\farcs0 slit, and one was
observed with a 0\farcs7 slit. A 1\farcs0 slit
leads to a spectral resolution of ${\sim}22,000$ at red wavelengths
and ${\sim}28,000$ at blue wavelengths. A 0\farcs7 slit yields a
spectral resolution of ${\sim}28,000$ and ${\sim}35,000$ in the red
and blue, respectively.
We used $2 \times 2$ on-chip binning to reduce read noise.
Stars were observed from ${\sim}6$ to $12$ hours, and individual frame
exposure times were typically ${\sim}55$ minutes to minimize cosmic
rays while collecting enough photons to avoid the read noise
limit. This was a problem particularly on the blue chip.
The signal-to-noise per pixel is modest, about 13 around
5300\,\AA\ and 26 around 6000\,\AA. 

Data were reduced with the CarPy MIKE python pipeline
\citep{Kelson03}\footnote{\url{http://code.obs.carnegiescience.edu/mike}}.
Consecutive observations were reduced together.
We normalized and stitched echelle orders before Doppler correcting
and combining spectra from different observation dates.
The Doppler correction was found by cross-correlation with a
template spectrum using the Mg triplet lines near 5200\,\AA.

\Booii\ was also observed at low resolution by \citet{Koch09}.
Our brightest star (\StarA) was the subject of \citet{Koch14b}. 
Of our other three target stars, two were identified as members of
\Booii\ in \citet{Koch09} (\StarB\ and \StarD). 

\begin{deluxetable*}{llllrrcccc}
\tablecolumns{10}
\tablewidth{0pt}
\tabletypesize{\scriptsize}
\tablecaption{Observing Details \label{tbl:obs}}
\tablehead{\colhead{Star} & \colhead{$\alpha$} & \colhead{$\delta$} &
  \colhead{Date of Observation} & \colhead{Slit} & \colhead{$t_{\rm exp}$} &
  \colhead{$g$} & \colhead{$V$\tablenotemark{a}} & \colhead{S/N\tablenotemark{b}} & \colhead{S/N\tablenotemark{b}}
\\
\colhead{} & \colhead{(J2000)} & \colhead{(J2000)} & \colhead{} &
\colhead{width} & \colhead{(hr)} & \colhead{(mag)} & \colhead{(mag)} &
\colhead{(5300\,\AA)} & \colhead{(6000\,\AA)}}
\startdata
\StarA & 13 57 51.2 & +12 51 37.0 & 2014 Mar 21, 2015 Jun 17/18 & 1\farcs0 & 5.8 & 19.24 & 18.86 & 11 & 21 \\
\StarB & 13 57 59.7 & +12 54 26.4 & 2014 Mar 11/12 & 1\farcs0 & 7.7 & 19.35 & 18.93 & 15 & 28 \\
\StarC & 13 57 56.2 & +12 52 07.7 & 2011 Mar 10/13, 2014 Mar 10/11 & 0\farcs7 & 9.2 & 19.56 & 19.15 & 17 & 29 \\
\StarD & 13 58 01.4 & +12 51 05.2 & 2010 Mar 18/19/21/22 & 1\farcs0 & 12.2 & 19.57 & 19.19 & 12 & 26
\enddata
\tablenotetext{a}{Converted from SDSS photometry using \citet{Schlafly11} and \citet{Jordi06}.}
\tablenotetext{b}{Signal-to-noise is per pixel.}
\end{deluxetable*}

\section{Chemical Abundances}\label{s:abund}
\subsection{Abundance Analysis}\label{s:abundmethods}
We used analysis software from \citet{Casey14} to measure equivalent widths,
determine stellar parameters, and obtain chemical abundances.
We use the \citet{Castelli04} 1D plane-parallel model atmospheres with
$\alpha$-enhancement and the LTE abundance analysis code MOOG
\citep{Sneden73} that accounts for Rayleigh scattering \citep{Sobeck11}. 
Final abundance ratios $\mbox{[X/Fe]}$ are relative to the
\citet{Asplund09} solar abundances\footnote{$\mbox{[X/Y]} = \log_{10}(N_X/N_Y) -
  \log_{10}(N_X/N_Y)_\odot$ for element X,Y}.

Equivalent widths of metal absorption lines were measured by fitting
Gaussian profiles to the line list from \citet{Roederer10}. The measurements are
given in Table~\ref{tbl:ew}.
We exclude lines whose reduced equivalent width ($\log (EW/\lambda)$)
is larger than $-4.5$, as such lines are likely past the linear regime
of the curve of growth.
Uncertainties on equivalent width measurements were calculated with
the equation from \citet{Frebel06a}, resulting in typical
uncertainties of ${\sim}14-18$\%.
For blended lines and the molecular CH features, we determined the
abundance with spectrum synthesis.

\input{ew_short.tex}

Stellar parameters are derived following the procedure in
\citet{Frebel13}, including the effective temperature correction.
Uncertainties in the stellar parameters are about 150K for $T_{\rm eff}$,
0.3 dex for \logg, and 0.15 \kms\ for \micro. 
The uncertainties for \StarA\ are slightly higher (200K, 0.4 dex,
0.2 \kms) due to the lower signal-to-noise.
The final stellar parameters are listed in Table~\ref{tbl:stellarparams}.
As an independent assessment of the effective temperature, we apply the
\citet{Alonso99} photometric $B-V$ temperature calibration assuming
$\mbox{[Fe/H]}=-3$. The $B-V$ color is found using SDSS $g$ and $r$,
the recalibrated reddening maps of \citet{Schlafly11}, and the
filter conversions of \citet{Jordi06}.
The two temperature determinations agree well within the uncertainty.

Random uncertainties in abundances are estimated from the dispersion in
individual line measurements. For elements with only one line, the
uncertainty is estimated by varying the placement of the continuum
corresponding to the uncertainty on the equivalent width.
A minimum abundance uncertainty of 0.1 dex is adopted for \StarB,
\StarC, and \StarD; and 0.15 dex for \StarA. 
For elements with no detectable lines, we determined an upper limit by
placing the continuum as high as reasonably possible around the
strongest line.
Systematic uncertainties due to differences in stellar parameters are
estimated by varying each parameter individually within its
uncertainties. Typical total uncertainties from stellar parameters are
$0.2-0.25$ dex, except for carbon which has a larger uncertainty of
${\sim}$0.4 dex due to its larger temperature sensitivity.

\begin{deluxetable}{cccccc}
\tablewidth{0pt}
\tabletypesize{\footnotesize}
\tabletypesize{\tiny}
\tablecaption{Stellar Parameters\label{tbl:stellarparams}}
\tablehead{\colhead{Star} & \colhead{$T_{\rm eff}$} & \colhead{\logg} &
  \colhead{\micro} & \colhead{$\mbox{[Fe/H]}$} & \colhead{$T_{\rm phot}$\tablenotemark{a}}\\
  \colhead{} & \colhead{(K)} & \colhead{} & \colhead{(\kms)} &
  \colhead{} & \colhead{(K)}}
\startdata
\StarA & 5102 & 2.20 & 2.00 & $-2.86$ & 5190 \\
\StarB & 4936 & 1.80 & 2.05 & $-2.63$ & 5005 \\
\StarC & 5013 & 2.00 & 2.00 & $-2.92$ & 5052 \\
\StarD & 5192 & 2.60 & 1.65 & $-2.87$ & 5217
\enddata
\tablenotetext{a}{Using converted SDSS magnitudes and $B-V$ calibration from \citet{Alonso99}}
\end{deluxetable}

\subsection{\Booii\ Abundance Signature}\label{s:booiiabund}
\begin{figure*}
  \begin{center}
    \includegraphics[width=16cm]{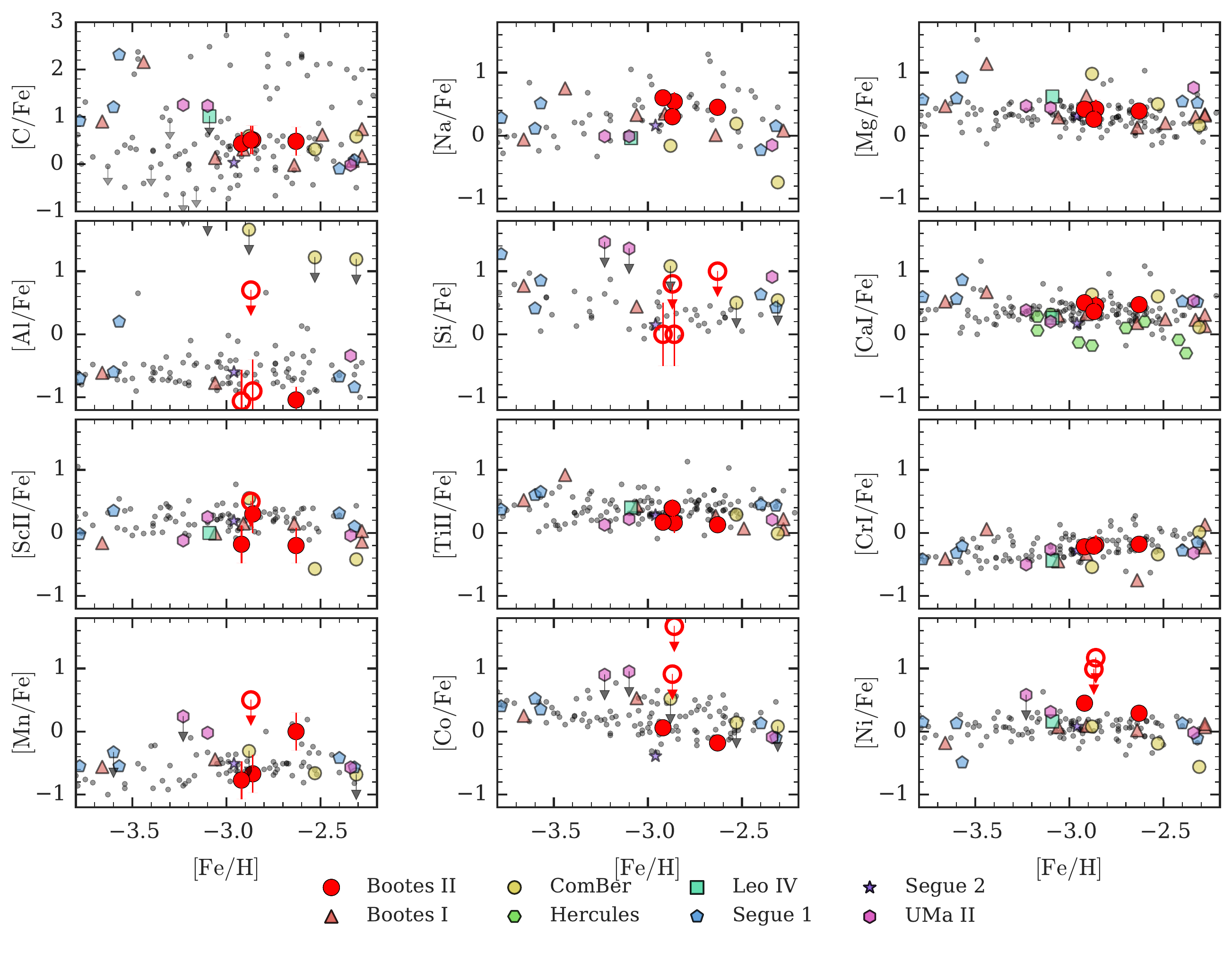}
  \end{center}
  \caption{\Booii\ chemical abundances compared to stars from the literature.
    Large red points are our \Booii\ measurements. 
    Small gray points are halo star abundances from \citet{Yong13a}. 
    Other colors are stars from other dwarf galaxies:
    Bo\"otes~1 (\citealt{Norris10a,Norris10b,Gilmore13,Ishigaki14}, Frebel et al (in prep)),
    ComBer \citep{Frebel10b},
    Hercules \citep{Koch08,Koch13}, Leo~IV \citep{Simon10}, 
    Segue~1 \citep{Frebel14}, Segue~2 \citep{Roederer14a},
    and UMa~II \citep{Frebel10b}.
    For stars in Bo\"otes~1 with more than one reference, we took
    abundances according to Frebel et al (in prep).
    All the y-axes are on the same scale except [C/Fe].
    Open \Booii\ points with error bars denote measurements where a
    nonzero amount of the element was detected in synthesis, but the
    abundances are very uncertain (see text).
    Upper limits are denoted by downward arrows.
  \label{f:abundances}}
\end{figure*}

\begin{figure*}
  \includegraphics[width=16cm]{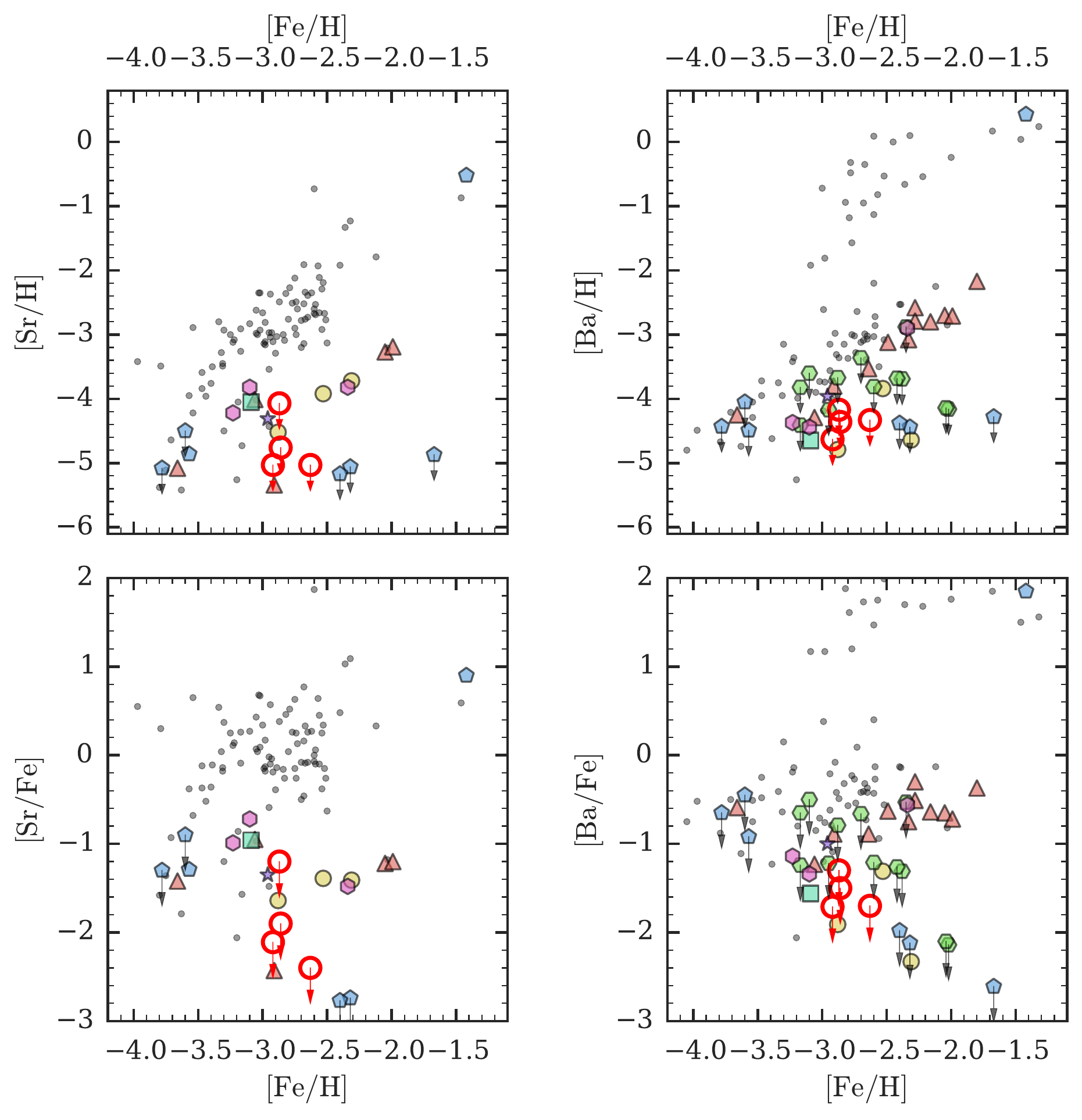}
  \caption{Neutron-capture elements Sr and Ba.
    Top row shows [X/H], bottom row shows [X/Fe].
    Symbols are the same as Figure~\ref{f:abundances}.
    The \Booii\ abundances are all upper limits. With the exception of
    one star in Segue~1 that is likely the recipient of mass transfer
    \citep{Frebel14}, most stars in faint dwarf galaxies have low
    neutron-capture elements compared to the halo.
  \label{f:ncap}}
\end{figure*}

We now discuss the individual element abundances of our \Booii\ stars.
We measure the abundances of C, Na, Mg, Al, Si, Ca, Sc, Ti, Cr, Mn,
Fe, Co, Ni, Sr, and Ba.
Unless otherwise indicated, abundances were obtained from equivalent
width measurements.
The abundances of all four stars are listed in Table~\ref{tbl:abund}
and plotted against halo stars and other ultra-faint dwarf galaxy
stars in Figures~\ref{f:abundances}~and~\ref{f:ncap}.
As a representative halo star sample, we use the homogeneous abundance
analysis of 190 halo stars from \citet{Yong13a}.
Medium-resolution spectroscopy abundance determinations of stars in
UFDs are consistent with the high-resolution spectroscopy abundances,
though with larger error bars \citep{Vargas13}.

\subsubsection{Carbon}
The carbon abundance of stars in dwarf galaxies may be an important
tracer of nucleosynthesis in the first stars (e.g.,
\citealt{Norris13,Cooke14,Salvadori15,Ji15}). 
We measure the carbon abundance through synthesis of the CH molecular
bands at ${\sim}4313$\,\AA\ and 4323\,\AA\ and correct their abundance
for the stars' evolutionary status
\citep{Placco14}\footnote{\url{http://www3.nd.edu/~vplacco/carbon-cor.html}}.
None of the stars are Carbon Enhanced Metal Poor (CEMP) stars, as all
stars have $\mbox{[C/Fe]} < 0.7$ \citep{Aoki07}.
However, all four stars display a mild $\mbox{[C/Fe]}$ enhancement
(${\sim}0.4-0.5$) comparable to other stars in
UFDs (see Figure~\ref{f:abundances}).
Note that we have corrected the dwarf galaxy stars according to
\citet{Placco14}, but not the halo star sample. This correction is
small for our hot stars ($0.01-0.03$ dex), but larger for
the cool star \StarB\ ($0.19$ dex).

Besides its role in understanding Population~III nucleosynthesis,
carbon also plays a role in forming the first low-mass stars, where it
(along with oxygen) is an important contributor to atomic metal line
cooling in the early universe \citep{Bromm03}. A sufficient amount of
carbon and oxygen will trigger vigorous fragmentation in a primordial
galaxy, resulting in the formation of multiple stars and star clusters
\citep{Bromm01,SafShrad14a}. The critical amount of carbon and oxygen
is specified by the $D_{\rm trans}$ criterion, which quantifies the
minimum amount of C and O required for metal line cooling to overcome
heating from adiabatic contraction \citep{Frebel07}. 
Oxygen is not detectable in our spectra, but a carbon abundance of
$\mbox{[C/H]} > -3.5$ in all our stars is already sufficient to
satisfy the $D_{\rm trans}$ criterion and trigger
significant atomic line cooling, allowing atomic cooling to play an
important role in forming the first low-mass stars in \Booii.

\subsubsection{$\alpha$-elements: Mg, Si, Ca, Ti}
The $\alpha$-element abundances of metal-poor halo stars are generally
enhanced at a level of $\mbox{[$\alpha$/Fe]} \sim 0.4$, which is typical of
stars whose birth cloud was enriched only by core-collapse supernovae. 
We measure the $\alpha$-elements magnesium, silicon, calcium, and
titanium, of which $\mbox{[Mg/Fe]}$, $\mbox{[Ca/Fe]}$,
and $\mbox{[Ti/Fe]}$ are enhanced. This is completely consistent with
halo star abundances and suggests enrichment only by massive stars.

Due to its potential significance in understanding the role of dust
cooling during the formation of the first low-mass stars \citep{Cherchneff10,Ji14}, we have
attempted to obtain silicon measurements or upper limits for all of our
stars. Only the 3905\,\AA\ Si\,I line, which is blended with carbon,
is strong enough to be detected. However, this line is in a region of
our spectra with relatively low signal-to-noise.
For \StarB\ and \StarD, we set upper limits. The silicon
abundance in both \StarA\ and \StarC\ is clearly nonzero, but the
signal-to-noise in this region is insufficient to allow a good
determination of the abundance (i.e., uncertainties at least 0.5
dex). We mark these points as open circles with error bars in
Figure~\ref{f:abundances}. Adding 1.0 dex to the tabulated abundances
is also an appropriate upper limit. All silicon abundances and upper
limits are consistent with the standard halo pattern of
$\mbox{[Si/Fe]} \sim 0.4$.
None of these upper limits or measurements is low enough to exclude
silicate dust cooling in the formation of these low-mass stars
\citep{Ji14}.

\subsubsection{Iron-peak elements: Cr, Mn, Co, Ni}
The chromium abundances in all our stars conform well to the tight
$\mbox{[Cr/Fe]}$ relation found in halo stars. 
We determine the manganese abundance from synthesis of the triplet at
${\sim}$4030\,\AA. In \StarD, we detect no clear manganese lines and give a
conservative upper limit.
Co and Ni are detected with few lines only in the cooler two stars
(\StarB\ and \StarC), and we determined upper limits for the
remaining two stars.
Overall, the abundances of the iron-peak elements are consistent with
the abundances of halo stars.

\subsubsection{Odd elements: Na, Al, Sc}
Sodium abundances were determined from the Na D doublet. The
abundance derived from these lines usually has a large NLTE
correction, but corrections were not applied to our comparison halo
sample \citep{Yong13a}. Thus to compare to that sample, we do not
apply a correction to the abundances in Figure~\ref{f:abundances}, and only LTE
abundances are listed in Table~\ref{tbl:abund}.
For reference, we also determine the NLTE corrections using
\citet{Lind11}\footnote{\url{www.inspect-stars.com}} with our equivalent
widths and stellar parameters. The NLTE abundances are typically lower 
by $\sim 0.4$ dex.
The exact abundance differences for the 5889.95{\AA} line are $-$0.46,
$-$0.52; $-$0.57; and $-$0.39 dex; for the 5895.92{\AA} line are $-$0.43,
$-$0.46, $-$0.47, and $-$0.31 dex;
for \StarA, \StarB, \StarC, and \StarD, respectively.

We determine the aluminum abundance from two lines. The
3944\,\AA\ line is blended with carbon, requiring spectrum synthesis. 
As in the case of silicon, for \StarC\ and \StarA\ we detect
nonzero aluminum but cannot put good bounds on the error (open circles
with error bars in Figure~\ref{f:abundances}) due to low
signal-to-noise. Again, adding 1.0 dex to the abundance in
Table~\ref{tbl:abund} is an appropriate upper limit.

The scandium abundances are determined through synthesis
of up to five different lines. In all four stars, the abundances are
consistent with the halo, though biased slightly lower. 
Upon investigating this slight bias, we found that the Sc abundances from
\citet{Yong13a} were determined using the \texttt{blends} driver of
MOOG, rather than from synthesis.
When we determine the Sc abundance with \texttt{blends}, we find our derived
Sc abundance to be systematically $0.1-0.3$ dex higher. This
difference deserves further investigation, but since the uncertainty
on the scandium measurements in \Booii\ is large we do not consider it
further here.

\subsubsection{Neutron-capture elements: Sr, Ba}
None of the four stars have strontium or barium line detections (e.g.,
Figure~\ref{f:specoverview}). We
use the strongest lines to place upper limits, i.e. the 4077\,\AA\ line
for Sr and the 4554\,\AA\ line for Ba.
We find $\mbox{[Sr/H]} \lesssim -5$, except for \StarD\ which has
$\mbox{[Sr/H]} \lesssim -4$. The limit for Ba is $\mbox{[Ba/H]}
  \lesssim -4.5$.
Such a deficiency of neutron-capture elements has been observed in
other UFDs, including Hercules \citep{Koch13} and Segue~1
\citep{Frebel14}.
We discuss this more in Section~\ref{s:concl}.

\subsection{Comparison with \citet{Koch14b}}
\StarA\ has also been observed at high resolution by
\citet{Koch14b}, who number the star as \Booii-15.
For their stellar parameters, they obtain $T_{eff}=5000$\,K,
\logg$=2.26$, \micro$=1.81$\,\kms, and $\mbox{[Fe/H]}=-2.93$.
These parameters are consistent within 1$\sigma$ error bars of our
stellar parameters.
We adopt their stellar parameters but use our equivalent widths to do
an abundance comparison.
The new abundances agree with \citet{Koch14b} within the 1$\sigma$
uncertainties, except for chromium. 
Our chromium abundance of $\mbox{[Cr/Fe]}=-0.18$ is only slightly more
than 1$\sigma$ discrepant with their measurement of $-0.38$.

\input{abund.tex}

\section{A Binary Star: \StarA}\label{s:rv}
\begin{figure}
  \includegraphics[width=8cm]{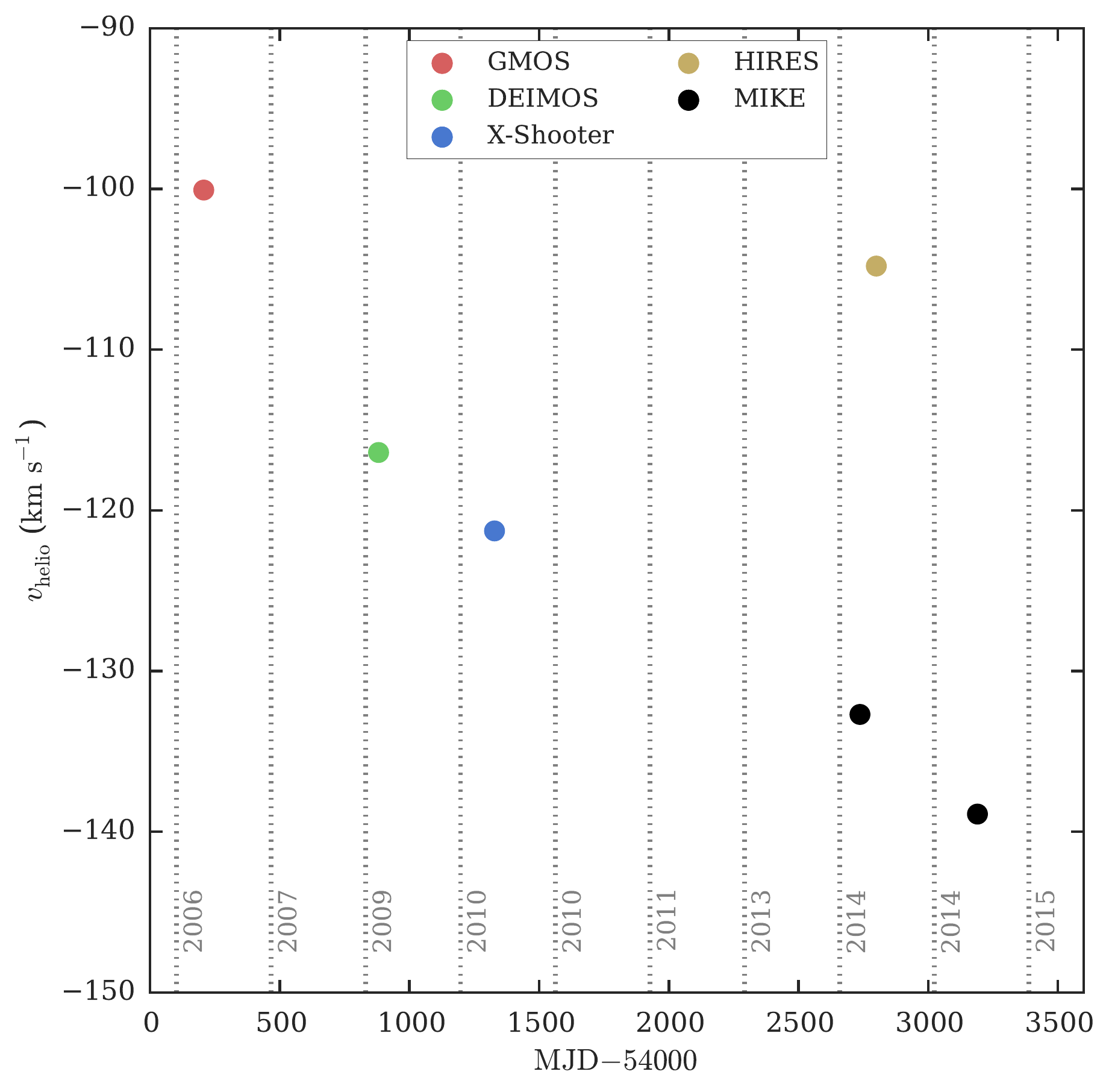}
  \caption{Heliocentric radial velocity over time. 
    From left to right:
    Red dot is Gemini/GMOS \citep{Koch09}, green dot is Keck/DEIMOS (\Gehaprep),
    blue dot is VLT/X-Shooter, yellow dot is Keck/HIRES
    \citep{Koch14b}, black dots are Magellen/MIKE (this work). Grey 
    lines denote Jan 1 of the labeled year.
  \label{f:rv}}
\end{figure}

The star \StarA\ (\Booii-15 in \citealt{Koch09,Koch14b}) has exhibited
significant variations in its heliocentric radial velocity with time.
We combine our radial velocity measurements with those from the
literature (\citealt{Koch09,Koch14b}, \Gehaprep) as well
as a publically available spectrum of this star taken with
VLT/X-Shooter. The velocity measurements are plotted in
Figure~\ref{f:rv}. Other than our two measurements from MIKE, we
note that every data point has been taken 
with a different instrument (the others are from Gemini/GMOS,
Keck/DEIMOS, VLT/X-Shooter, Keck/HIRES).
There may be systematic zero-point offsets from instrument to instrument
of the order a few \kms, but not enough to account for the
observed differences. In contrast, the velocities of the other two
stars observed with MIKE are consistent with \citet{Koch09} to
within $2$\,\kms.

This amount of velocity variation would be consistent with a RR Lyrae
star, but no line distortions (e.g., in the hydrogen lines) from
rapidly changing stellar parameters are seen in our MIKE spectra that
would be consistent with this type of variable star.
Our two observations are at a similar phase in the velocity
variations, but the stellar parameters from our study and
\citet{Koch14b} are consistent within errors despite being determined
at different very different phases of the velocity variations, while
an RR Lyrae's effective temperature changes by almost 1000\,K
\citep{Fossati14}.
Thus, we conclude it is likely that this star is in a binary system.
If so, it would be the third binary system confirmed in an ultra-faint
dwarf. A full binary orbit was established for a star in Hercules
\citep{Koch14a}, and one star in Segue~1 shows clear signs of a past
mass transfer event \citep{Frebel14}. One star in Ursa~Major~II may
also show a signature of binary mass transfer \citep{Frebel10b}.
We note that one of our MIKE observations and the HIRES observation from
\citet{Koch14b} are only two months apart, but the measurements differ
by ${\sim}30$ \kms. This rapid velocity shift is consistent with the
eccentric orbit seen in the Hercules system \citep{Koch14a}.
Given only six observations and no clear way to determine
velocity errors across different instruments, we were unable to
robustly determine a period, amplitude, or mean velocity for the
orbit.

The binarity of this system has had an adverse effect on the
mass determination of \Booii. The velocity dispersion in
\citet{Koch09} was determined to be $10.5 \pm 7.4$\,\kms, where the
large error bar is due to the low spectral resolution and only using
five stars to determine the 
dispersion. However, they observed \StarA\ to have a velocity of 
$-100.1$\,\kms, and this star is the most extreme value in their
velocity distribution. 
Although we cannot robustly determine the mean velocity of \StarA, it
is likely much closer to the systemic velocity of ${\sim}-120$\,\kms\
than the originally observed velocity of ${\sim}-100$\,\kms.
If we remove \StarA\ from their sample, the large errors on the
remaining four velocities result in only an upper limit on the
velocity dispersion for this system.
In comparison, \Gehaprep\ find a velocity dispersion in line with
other UFDs of similar size and luminosity.
We note that the large velocity dispersion was included as part of the 
commonly-used compilation of Local Group properties in
\citet{McConnachie12}, where as a result \Booii\ is listed as the
galaxy with the largest mass-to-light ratio currently known.
The discrepancy can affect works that use the \citet{McConnachie12} 
compilation (e.g., \citealt{Jiang15}).

\section{Discussion and Conclusion}\label{s:concl}
We have obtained high-resolution spectra of four metal-poor red giants
in the ultra-faint dwarf galaxy \Booii. Chemical abundance analysis
shows that all detectable elements have abundance ratios consistent
with metal-poor stars in the stellar halo \citep{Yong13a}, 
with the exception of Sr and Ba. The two neutron-capture elements have
very low abundance limits, consistent with what has been observed in
other UFDs \citep{Koch13,Frebel14}.
In addition, one of the stars exhibits significant radial velocity
variations and is likely a single-lined spectroscopic binary. The
binary system has led to an overestimation of the \Booii\ velocity
dispersion in the past (see Section~\ref{s:rv} for the resulting
consequences).
We now discuss some possible interpretations of the chemical abundance
pattern found in \Booii.

\emph{\Booii\ as a first galaxy}.
\citet{Frebel12} define a 
first galaxy as a galaxy that has experienced very
few metal enrichment events, perhaps only from Pop~III stars. They
suggest that some UFDs may be surviving first galaxies.
This scenario results in a wide metallicity spread ($\gtrsim 1$ dex),
super-solar $\alpha$-abundances at all metallicities, and low
neutron-capture abundances. Based on these criteria, Segue~1 is
potentially a surviving first galaxy
\citep{Frebel14}, and we now consider whether \Booii\ has a
chemical signature consistent with being a first galaxy.

The overall metallicity distribution function of \Booii\ 
is wide, similar to the other UFDs (\Gehaprep). However, our
high-resolution abundances span only the metal-poor end of the
metallicity distribution. Thus, although we find enhanced
$\alpha$-abundances in all our stars, the limited metallicity range is
insufficient to conclusively determine that \Booii\ shows the enhanced
$\alpha$ signature across its entire metallicity range.
If the high $\alpha$-abundances seen at low metallicity in \Booii\
extend to the higher metallicity stars, this is a strong indication
that star formation in \Booii\ ended before the onset of Type~Ia
supernovae, suggesting \Booii\ is a candidate first galaxy.
In contrast, a downturn in the $\mbox{[$\alpha$/Fe]}$ ratio at higher
metallicities would suggest that \Booii\ experienced some extended
but inefficient star formation.
Unfortunately, the higher-metallicity stars in \Booii\ are all too
faint to be observed with high-resolution spectrographs on current
telescopes. Medium-resolution spectra may be able to constrain the
$\alpha$-abundances of these fainter stars \citep{Vargas13}.
Either way, the low neutron-capture abundances in \Booii\ stars are
clearly lower than the pattern established by halo stars and likely
indicates these stars in \Booii\ formed before standard
neutron-capture enrichment processes began to enrich the system.

\emph{Source of neutron-capture elements in UFDs.}
Although we only have upper limits on the neutron-capture abundances,
\citet{Roederer13} points out that no stars have yet been observed
with enough signal-to-noise to clearly rule out the presence of
neutron-capture elements.
In addition, very low but non-zero neutron-capture abundances have been
detected in many other UFDs.
Supposing that \Booii\ and other dwarf galaxies contain nonzero but
small amounts of neutron-capture elements, it is interesting to
consider the potential source of these heavy elements.
Following \citet{Frebel14}, we can use the current upper limits to
estimate the maximum mass of neutron-capture material in a galaxy like
\Booii\ by assuming it formed from $10^5\,M_\odot$ of gas. Then the
upper limits $\mbox{[Sr/H]}<-5$ and $\mbox{[Ba/H]}<-4.5$ suggest the
mass of each element is $\lesssim 10^{-7}\,M_\odot$. If the initial gas
reservoir is larger, the mass limit scales accordingly
(e.g., a $10^7\,M_\odot$ gas reservoir puts the limit at $\lesssim
10^{-5}\,M_\odot$).

The standard source of both Sr and Ba is the stellar winds from
intermediate mass AGB stars (see references in \citealt{Jacobson14}).
From \citet{Lugaro12}, we estimate that a typical intermediate-mass
AGB star releases ${\sim}1\,M_\odot$ of stellar winds with
$\mbox{[Sr/Fe]}\sim 1.5$ and $\mbox{[Ba/Fe]}\sim 2.0$. Assuming the
star has $\mbox{[Fe/H]}=-3$, this results in
${\sim}10^{-8.5}\,M_\odot$ of Sr and Ba per AGB star. 
The neutron-capture mass constraint in \Booii\ allows
$\lesssim 30$ Pop~II AGB stars to enrich the system.
However, we believe standard AGB enrichment to be unlikely in
these systems, because the $\mbox{[Sr,Ba/Fe]}$ ratios are distinct from
the halo stars. This suggests the dominant source of neutron-capture
elements in halo stars is different than in UFDs.
We thus look to other explanations for the low neutron-capture
abundances in UFDs.

One possibility is that this low neutron-capture abundance signature
indicates unique nucleosynthesis processes in the first (Pop~III)
stars. For example, massive rapidly-rotating Pop~III stars could
produce Sr and Ba masses around $10^{-9}\,M_\odot$ per star through
a special s-process \citep{Frischknecht12}.
However, this amount depends heavily on the presence of seed nucleii, and
these models typically invoke a small initial metallicity in their
star to seed the s-process.

Alternatively, Pop~III core-collapse supernovae (CCSNe) can produce
neutron-capture elements if they undergo the r-process, a possibility
that has support from abundances of metal-poor halo stars \citep{Roederer14b}.
Note that theoretically, the conditions for producing a realistic
r-process pattern in CCSNe are not easily achieved, leading many to
look to merging neutron star binaries as the primary source of the
r-process (e.g., \citealt{Lattimer76,Freiburghaus99,Korobkin12}). 
While this pathway may be important for higher metallicity stars, the
rarity of these binaries and the long merging timescales suggests they
are not important at low metallicities (\citealt{Argast04}, although
see \citealt{Shen15}). 
In addition, a single compact binary merger typically produces a
neutron-capture mass of $10^{-3}-10^{-2}\,M_\odot$ \citep{Goriely11},
which is far above the observed $10^{-7}-10^{-5}\,M_\odot$ limit in UFDs.
Thus, it appears CCSNe may be the most likely source of detected
neutron-capture elements in UFDs discovered so far.

\citet{Lee13} suggest that the difference between halo and UFD
neutron-capture abundance distributions is a stochastic effect due to a small number
of stars enriching UFDs combined with a strongly mass-dependent
yield. This would imply that some UFDs should have members 
with neutron-capture abundance ratios above the halo mean. This has
not been seen yet even though the number of stars observed with
high-resolution spectroscopy in UFDs has
roughly doubled compared to two years ago. Still, these additional data may
not be enough to exclude the possibility of mass-dependent yields.
However, there is also evidence that the Initial Mass Function (IMF)
in UFDs is top-heavy \citep{Geha13}. If neutron-capture elements are
preferentially produced in $8-10\,M_\odot$ core-collapse supernovae
(e.g., \citealt{Wanajo03}) then a biased IMF could suppress neutron-capture
abundances in UFDs. Note that the galaxy would have to undergo at
least one episode of self-enrichment for this bias to manifest itself
in the surviving stars.

\emph{Signatures of first stars.} 
As they are likely enriched by very few generations of stars, UFDs
are a promising place to look for chemical signatures from the 
first stars \citep{Ji15}. 
The CEMP signature in particular has attracted a lot of attention,
with Pop~III stars emerging as a strong candidate to produce these
strange abundance signatures (see \citealt{Norris13} for a
comprehensive overview).
The four stars in \Booii\ have mild carbon overabundances
($\mbox{[C/Fe]} \sim 0.5$), but not beyond the CEMP threshold
of $\mbox{[C/Fe]} > 0.7$. 
Compared to the halo stars, this is not too surprising: only
${\sim}35\%$ of halo stars with $\mbox{[Fe/H]}\leq-3$ are CEMP stars,
and thus the probability that four stars with $\mbox{[Fe/H]} \sim -3$
are all not CEMP is ${\sim}18\%$ \citep{Placco14}.
In addition, when comparing to other UFD stars, stars with
$\mbox{[Fe/H]} \gtrsim -3$ have at most mild carbon enhancements. It
is only stars with $\mbox{[Fe/H]} < -3$ that begin showing large
$\mbox{[C/Fe]}$ values.
If most or all Pop~III stars produce carbon-enhanced metal yields
(e.g., \citealt{Salvadori15}), this would suggest that the four stars
we observed have additionally been enriched by some standard
core-collapse supernovae. These could be either Pop~III or Pop~II
supernovae, and a better understanding of \Booii's metal enrichment
history is needed to decide whether \Booii\ is a sufficiently
unenriched system to cleanly study Pop~III yields.

While UFD abundance patterns have traditionally been compared to
abundances of Milky Way halo stars,
there are now enough UFDs (defined as dwarf galaxies around the Milky
Way with $M_V > -7$) with chemical abundances to begin establishing
their own abundance pattern.
The halo is a complicated amalgam of stars
from several different sources (e.g., small satellites, large
satellites, heated disk stars, etc.), and the faintest dwarfs may be
a more homogeneous population all probing a more specific time in
cosmic history.
\Booii\ exemplifies the emerging UFD abundance pattern, which
is similar to the halo in all elements except the particularly low
neutron-capture abundances. The chemical signature in more and more
UFDs suggests they are clear candidates to be ancient survivors from
the beginning of the universe, consistent with their star-formation
histories \citep{Brown14}.

\acknowledgements
We thank Vini Placco for assistance reducing the X-Shooter spectra,
and Heather Jacobson for investigating the scandium abundances.
APJ and AF are supported by NSF-CAREER grant AST-1255160. AF
acknowledges support from the Silverman (1968) Family Career
Development Professorship.
JDS acknowledges support from grant AST-1108811.
This work made extensive use of NASA's Astrophysics Data System
Bibliographic Services and the python libraries \texttt{numpy},
\texttt{scipy}, \texttt{matplotlib}, and \texttt{seaborn}.

\end{document}

%% file: ew_short.tex
\begin{deluxetable*}{llrrrrrrrrrr}
\tablecolumns{12}
\tablewidth{0pt}
\tabletypesize{\footnotesize}
\tabletypesize{\tiny}
\tablecaption{Equivalent Widths\label{tbl:ew}}
\tablehead{\colhead{El.} & \colhead{$\lambda$} & \colhead{$\chi$} & \colhead{$\log gf$} & \colhead{EW (m\AA)} & \colhead{$\log \epsilon$ (dex)} & \colhead{EW (m\AA)} & \colhead{$\log \epsilon$ (dex)} & \colhead{EW (m\AA)} & \colhead{$\log \epsilon$ (dex)} & \colhead{EW (m\AA)} & \colhead{$\log \epsilon$ (dex)}\\
\colhead{} & \colhead{(\AA)} & \colhead{(eV)} & \colhead{(dex)} & \multicolumn{2}{c}{\StarA} & \multicolumn{2}{c}{\StarB} & \multicolumn{2}{c}{\StarC} & \multicolumn{2}{c}{\StarD}}
\startdata
CH     & 4313     &\nodata&\nodata   &     syn &    5.97 &     syn &    6.10 &     syn &    5.91 &     syn &    6.06 \\
CH     & 4323     &\nodata&\nodata   &     syn &    5.97 &     syn &    6.15 &     syn &    5.91 &     syn &    6.06 \\
Na\,I  &  5889.95 &  0.00 & $  0.11$ &   151.4 &    3.84 &   186.0 &    4.09 &   168.2 &    3.96 &   133.4 &    3.68 \\
Na\,I  &  5895.92 &  0.00 & $ -0.19$ &   142.7 &    4.00 &   160.3 &    4.04 &   144.2 &    3.89 &   114.5 &    3.66 \\
\enddata
\tablecomments{This table is available in its entirety in a machine-readable form in the online journal and the arXiv source. A portion is shown here for guidance regarding its form and content.}
\end{deluxetable*}

%% file: abund.tex
\begin{deluxetable}{lrrcrr}
\tablecolumns{6}
\tablewidth{0pt}
\tabletypesize{\footnotesize}
\tabletypesize{\tiny}
\tablecaption{\label{tbl:abund}}
\tablehead{\colhead{Species} & \colhead{$N$} & \colhead{$\log \epsilon(X)$} & $\sigma$ & $\mbox{[X/H]}$ & $\mbox{[X/Fe]}$}
\hline
\startdata
\cutinhead{ \StarA }
CH         &   2 &     5.97 &    0.30 &  $-$2.46 &   0.41 \\
Na I       &   2 &     3.92 &    0.15 &  $-$2.32 &   0.54 \\
Mg I       &   4 &     5.16 &    0.15 &  $-$2.44 &   0.42 \\
Al I       &   2 &     2.69:&    0.50 &  $-$3.76:&$-$0.90:\\
Si I       &   1 &     4.65:&    0.50 &  $-$2.86:&   0.00:\\
Ca I       &   4 &     3.93 &    0.15 &  $-$2.41 &   0.45 \\
Sc II      &   3 &     0.59 &    0.30 &  $-$2.56 &   0.30 \\
Ti II      &  11 &     2.25 &    0.15 &  $-$2.70 &   0.16 \\
Cr I       &   4 &     2.59 &    0.15 &  $-$3.05 &$-$0.19 \\
Mn I       &   3 &     1.90 &    0.30 &  $-$3.53 &$-$0.67 \\
Fe I       &  60 &     4.64 &    0.20 &  $-$2.86 &   0.00 \\
Fe II      &   7 &     4.66 &    0.15 &  $-$2.84 &   0.02 \\
Co I       &   2 &  $<$3.81 & \nodata & $<-$1.19 &  $<$1.68 \\
Ni I       &   1 &  $<$4.53 & \nodata & $<-$1.69 &  $<$1.17 \\
Sr II      &   1 & $<-$1.89 & \nodata & $<-$4.76 & $<-$1.90 \\
Ba II      &   1 & $<-$2.18 & \nodata & $<-$4.36 & $<-$1.50 \\
\cutinhead{ \StarB }
CH         &   2 &     6.12 &    0.20 &  $-$2.30 &   0.51 \\
Na I       &   2 &     4.06 &    0.10 &  $-$2.18 &   0.45 \\
Mg I       &   5 &     5.36 &    0.12 &  $-$2.24 &   0.39 \\
Al I       &   2 &     2.78 &    0.31 &  $-$3.67 &$-$1.18 \\
Si I       &   1 &  $<$5.88 & \nodata & $<-$1.63 &  $<$1.00 \\
Ca I       &   8 &     4.18 &    0.10 &  $-$2.16 &   0.47 \\
Sc II      &   5 &     0.32 &    0.28 &  $-$2.83 &$-$0.20 \\
Ti I       &   2 &     2.69 &    0.10 &  $-$2.27 &   0.36 \\
Ti II      &  17 &     2.45 &    0.11 &  $-$2.50 &   0.13 \\
Cr I       &   6 &     2.83 &    0.10 &  $-$2.81 &$-$0.18 \\
Mn I       &   3 &     2.80 &    0.30 &  $-$2.63 &$-$0.00 \\
Fe I       & 109 &     4.87 &    0.13 &  $-$2.63 &   0.00 \\
Fe II      &   9 &     4.89 &    0.10 &  $-$2.61 &   0.02 \\
Co I       &   3 &     2.18 &    0.10 &  $-$2.81 &$-$0.18 \\
Ni I       &   3 &     3.88 &    0.12 &  $-$2.34 &   0.29 \\
Sr II      &   1 & $<-$2.16 & \nodata & $<-$5.03 & $<-$2.40 \\
Ba II      &   1 & $<-$2.15 & \nodata & $<-$4.33 & $<-$1.70 \\
\cutinhead{ \StarC }
CH         &   2 &     5.91 &    0.25 &  $-$2.49 &   0.43 \\
Na I       &   2 &     3.92 &    0.10 &  $-$2.32 &   0.61 \\
Mg I       &   5 &     5.10 &    0.12 &  $-$2.50 &   0.42 \\
Al I       &   2 &     2.47:&    0.50 &  $-$3.98:&$-$1.05:\\
Si I       &   1 &     4.59:&    0.50 &  $-$2.92:&   0.00:\\
Ca I       &   6 &     3.92 &    0.10 &  $-$2.42 &   0.50 \\
Sc II      &   3 &     0.05 &    0.30 &  $-$3.10 &$-$0.18 \\
Ti II      &  12 &     2.20 &    0.10 &  $-$2.75 &   0.17 \\
Cr I       &   4 &     2.50 &    0.10 &  $-$3.14 &$-$0.21 \\
Mn I       &   3 &     1.74 &    0.27 &  $-$3.69 &$-$0.76 \\
Fe I       &  71 &     4.58 &    0.17 &  $-$2.92 &   0.00 \\
Fe II      &   5 &     4.56 &    0.10 &  $-$2.94 &$-$0.02 \\
Co I       &   2 &     2.13 &    0.10 &  $-$2.86 &   0.06 \\
Ni I       &   2 &     3.75 &    0.10 &  $-$2.47 &   0.45 \\
Sr II      &   1 & $<-$2.05 & \nodata & $<-$4.92 & $<-$2.00 \\
Ba II      &   1 & $<-$2.24 & \nodata & $<-$4.42 & $<-$1.50 \\
\cutinhead{ \StarD }
CH         &   2 &     6.06 &    0.30 &  $-$2.37 &   0.51 \\
Na I       &   2 &     3.67 &    0.10 &  $-$2.57 &   0.30 \\
Mg I       &   4 &     4.99 &    0.10 &  $-$2.61 &   0.26 \\
Al I       &   1 &  $<$4.28 & \nodata & $<-$2.17 &  $<$0.70 \\
Si I       &   1 &  $<$5.64 & \nodata & $<-$1.87 &  $<$1.00 \\
Ca I       &   2 &     3.83 &    0.10 &  $-$2.51 &   0.35 \\
Sc II      &   3 &  $<$0.78 & \nodata & $<-$2.37 &  $<$0.50 \\
Ti II      &  11 &     2.47 &    0.14 &  $-$2.48 &   0.39 \\
Cr I       &   3 &     2.56 &    0.11 &  $-$3.08 &$-$0.21 \\
Mn I       &   3 &  $<$3.06 & \nodata & $<-$2.37 &  $<$0.50 \\
Fe I       &  42 &     4.63 &    0.11 &  $-$2.87 &   0.00 \\
Fe II      &   3 &     4.63 &    0.10 &  $-$2.87 &   0.00 \\
Co I       &   1 &  $<$3.03 & \nodata & $<-$1.96 &  $<$0.91 \\
Ni I       &   1 &  $<$4.34 & \nodata & $<-$1.88 &  $<$0.99 \\
Sr II      &   1 & $<-$1.20 & \nodata & $<-$4.07 & $<-$1.20 \\
Ba II      &   1 & $<-$1.99 & \nodata & $<-$4.17 & $<-$1.30 \\
\enddata
\tablecomments{: Denotes highly uncertain abundance (see text for details)}
\end{deluxetable}

%% file: booii.bbl
\begin{thebibliography}{}
\expandafter\ifx\csname natexlab\endcsname\relax\def\natexlab#1{#1}\fi

\bibitem[{{Alam} {et~al.}(2015){Alam}, {Albareti}, {Allende Prieto}, {Anders},
  {Anderson}, {Anderton}, {Andrews}, {Armengaud}, {Aubourg}, {Bailey}, \&
  et~al.}]{SDSSDR12}
{Alam}, S., {Albareti}, F.~D., {Allende Prieto}, C., {et~al.} 2015, \apjs, 219,
  12

\bibitem[{{Alonso} {et~al.}(1999){Alonso}, {Arribas}, \&
  {Mart{\'{\i}}nez-Roger}}]{Alonso99}
{Alonso}, A., {Arribas}, S., \& {Mart{\'{\i}}nez-Roger}, C. 1999, \aaps, 140,
  261

\bibitem[{{Aoki} {et~al.}(2007){Aoki}, {Beers}, {Christlieb}, {Norris}, {Ryan},
  \& {Tsangarides}}]{Aoki07}
{Aoki}, W., {Beers}, T.~C., {Christlieb}, N., {et~al.} 2007, \apj, 655, 492

\bibitem[{{Argast} {et~al.}(2004){Argast}, {Samland}, {Thielemann}, \&
  {Qian}}]{Argast04}
{Argast}, D., {Samland}, M., {Thielemann}, F.-K., \& {Qian}, Y.-Z. 2004, \aap,
  416, 997

\bibitem[{{Asplund} {et~al.}(2009){Asplund}, {Grevesse}, {Sauval}, \&
  {Scott}}]{Asplund09}
{Asplund}, M., {Grevesse}, N., {Sauval}, A.~J., \& {Scott}, P. 2009, \araa, 47,
  481

\bibitem[{{Bernstein} {et~al.}(2003){Bernstein}, {Shectman}, {Gunnels},
  {Mochnacki}, \& {Athey}}]{Bernstein03}
{Bernstein}, R., {Shectman}, S.~A., {Gunnels}, S.~M., {Mochnacki}, S., \&
  {Athey}, A.~E. 2003, Proc. SPIE, 4841, 1694

\bibitem[{{Bromm} {et~al.}(2001){Bromm}, {Ferrara}, {Coppi}, \&
  {Larson}}]{Bromm01}
{Bromm}, V., {Ferrara}, A., {Coppi}, P.~S., \& {Larson}, R.~B. 2001, \mnras,
  328, 969

\bibitem[{{Bromm} \& {Loeb}(2003)}]{Bromm03}
{Bromm}, V., \& {Loeb}, A. 2003, \nat, 425, 812

\bibitem[{{Brown} {et~al.}(2014){Brown}, {Tumlinson}, {Geha}, {Simon},
  {Vargas}, {VandenBerg}, {Kirby}, {Kalirai}, {Avila}, {Gennaro}, {Ferguson},
  {Mu{\~n}oz}, {Guhathakurta}, \& {Renzini}}]{Brown14}
{Brown}, T.~M., {Tumlinson}, J., {Geha}, M., {et~al.} 2014, \apj, 796, 91

\bibitem[{{Casey}(2014)}]{Casey14}
{Casey}, A.~R. 2014, ArXiv, arXiv:1405.5968

\bibitem[{{Castelli} \& {Kurucz}(2004)}]{Castelli04}
{Castelli}, F., \& {Kurucz}, R.~L. 2004, ArXiv, astro-ph/0405087

\bibitem[{{Cherchneff} \& {Dwek}(2010)}]{Cherchneff10}
{Cherchneff}, I., \& {Dwek}, E. 2010, \apj, 713, 1

\bibitem[{{Clem} {et~al.}(2008){Clem}, {Vanden Berg}, \& {Stetson}}]{Clem08}
{Clem}, J.~L., {Vanden Berg}, D.~A., \& {Stetson}, P.~B. 2008, \aj, 135, 682

\bibitem[{{Cooke} \& {Madau}(2014)}]{Cooke14}
{Cooke}, R.~J., \& {Madau}, P. 2014, \apj, 791, 116

\bibitem[{{Drlica-Wagner} {et~al.}(2015){Drlica-Wagner}, {Albert}, {Bechtol},
  {Wood}, {Strigari}, {S{\'a}nchez-Conde}, {Baldini}, {Essig}, {Cohen-Tanugi},
  {Anderson}, {Bellazzini}, {Bloom}, {Caputo}, {Cecchi}, {Charles}, {Chiang},
  {de Angelis}, {Funk}, {Fusco}, {Gargano}, {Giglietto}, {Giordano}, {Guiriec},
  {Gustafsson}, {Kuss}, {Loparco}, {Lubrano}, {Mirabal}, {Mizuno}, {Morselli},
  {Ohsugi}, {Orlando}, {Persic}, {Rain{\`o}}, {Sehgal}, {Spada}, {Suson},
  {Zaharijas}, {Zimmer}, {The Fermi-LAT Collaboration}, {Abbott}, {Allam},
  {Balbinot}, {Bauer}, {Benoit-L{\'e}vy}, {Bernstein}, {Bernstein}, {Bertin},
  {Brooks}, {Buckley-Geer}, {Burke}, {Carnero Rosell}, {Castander},
  {Covarrubias}, {D'Andrea}, {da Costa}, {DePoy}, {Desai}, {Diehl}, {Cunha},
  {Eifler}, {Estrada}, {Evrard}, {Fausti Neto}, {Fernandez}, {Finley},
  {Flaugher}, {Frieman}, {Gaztanaga}, {Gerdes}, {Gruen}, {Gruendl},
  {Gutierrez}, {Honscheid}, {Jain}, {James}, {Jeltema}, {Kent}, {Kron},
  {Kuehn}, {Kuropatkin}, {Lahav}, {Li}, {Luque}, {Maia}, {Makler}, {March},
  {Marshall}, {Martini}, {Merritt}, {Miller}, {Miquel}, {Mohr}, {Neilsen},
  {Nord}, {Ogando}, {Peoples}, {Petravick}, {Pieres}, {Plazas}, {Queiroz},
  {Romer}, {Roodman}, {Rykoff}, {Sako}, {Sanchez}, {Santiago}, {Scarpine},
  {Schubnell}, {Sevilla}, {Smith}, {Soares-Santos}, {Sobreira}, {Suchyta},
  {Swanson}, {Tarle}, {Thaler}, {Thomas}, {Tucker}, {Walker}, {Wechsler},
  {Wester}, {Williams}, {Yanny}, {Zuntz}, \& {The DES Collaboration}}]{DrWag15}
{Drlica-Wagner}, A., {Albert}, A., {Bechtol}, K., {et~al.} 2015, \apjl, 809, L4

\bibitem[{{Fossati} {et~al.}(2014){Fossati}, {Kolenberg}, {Shulyak}, {Elmasli},
  {Tsymbal}, {Barnes}, {Guggenberger}, \& {Kochukhov}}]{Fossati14}
{Fossati}, L., {Kolenberg}, K., {Shulyak}, D.~V., {et~al.} 2014, \mnras, 445,
  4094

\bibitem[{{Frebel} \& {Bromm}(2012)}]{Frebel12}
{Frebel}, A., \& {Bromm}, V. 2012, \apj, 759, 115

\bibitem[{{Frebel} {et~al.}(2013){Frebel}, {Casey}, {Jacobson}, \&
  {Yu}}]{Frebel13}
{Frebel}, A., {Casey}, A.~R., {Jacobson}, H.~R., \& {Yu}, Q. 2013, \apj, 769,
  57

\bibitem[{{Frebel} {et~al.}(2006){Frebel}, {Christlieb}, {Norris}, {Aoki}, \&
  {Asplund}}]{Frebel06a}
{Frebel}, A., {Christlieb}, N., {Norris}, J.~E., {Aoki}, W., \& {Asplund}, M.
  2006, \apjl, 638, L17

\bibitem[{{Frebel} {et~al.}(2007){Frebel}, {Johnson}, \& {Bromm}}]{Frebel07}
{Frebel}, A., {Johnson}, J.~L., \& {Bromm}, V. 2007, \mnras, 380, L40

\bibitem[{{Frebel} {et~al.}(2010{\natexlab{a}}){Frebel}, {Kirby}, \&
  {Simon}}]{Frebel10a}
{Frebel}, A., {Kirby}, E.~N., \& {Simon}, J.~D. 2010{\natexlab{a}}, \nat, 464,
  72

\bibitem[{{Frebel} \& {Norris}(2015)}]{Frebel15}
{Frebel}, A., \& {Norris}, J.~E. 2015, \araa, 53, 631

\bibitem[{{Frebel} {et~al.}(2010{\natexlab{b}}){Frebel}, {Simon}, {Geha}, \&
  {Willman}}]{Frebel10b}
{Frebel}, A., {Simon}, J.~D., {Geha}, M., \& {Willman}, B. 2010{\natexlab{b}},
  \apj, 708, 560

\bibitem[{{Frebel} {et~al.}(2014){Frebel}, {Simon}, \& {Kirby}}]{Frebel14}
{Frebel}, A., {Simon}, J.~D., \& {Kirby}, E.~N. 2014, \apj, 786, 74

\bibitem[{{Freiburghaus} {et~al.}(1999){Freiburghaus}, {Rosswog}, \&
  {Thielemann}}]{Freiburghaus99}
{Freiburghaus}, C., {Rosswog}, S., \& {Thielemann}, F.-K. 1999, \apjl, 525,
  L121

\bibitem[{{Frischknecht} {et~al.}(2012){Frischknecht}, {Hirschi}, \&
  {Thielemann}}]{Frischknecht12}
{Frischknecht}, U., {Hirschi}, R., \& {Thielemann}, F.-K. 2012, \aap, 538, L2

\bibitem[{{Geha} {et~al.}(2013){Geha}, {Brown}, {Tumlinson}, {Kalirai},
  {Simon}, {Kirby}, {VandenBerg}, {Mu{\~n}oz}, {Avila}, {Guhathakurta}, \&
  {Ferguson}}]{Geha13}
{Geha}, M., {Brown}, T.~M., {Tumlinson}, J., {et~al.} 2013, \apj, 771, 29

\bibitem[{{Gilmore} {et~al.}(2013){Gilmore}, {Norris}, {Monaco}, {Yong},
  {Wyse}, \& {Geisler}}]{Gilmore13}
{Gilmore}, G., {Norris}, J.~E., {Monaco}, L., {et~al.} 2013, \apj, 763, 61

\bibitem[{{Goriely} {et~al.}(2011){Goriely}, {Bauswein}, \&
  {Janka}}]{Goriely11}
{Goriely}, S., {Bauswein}, A., \& {Janka}, H.-T. 2011, \apjl, 738, L32

\bibitem[{{Gustafsson} {et~al.}(2008){Gustafsson}, {Edvardsson}, {Eriksson},
  {J{\o}rgensen}, {Nordlund}, \& {Plez}}]{Gustafsson08}
{Gustafsson}, B., {Edvardsson}, B., {Eriksson}, K., {et~al.} 2008, \aap, 486,
  951

\bibitem[{{Ishigaki} {et~al.}(2014){Ishigaki}, {Aoki}, {Arimoto}, \&
  {Okamoto}}]{Ishigaki14}
{Ishigaki}, M.~N., {Aoki}, W., {Arimoto}, N., \& {Okamoto}, S. 2014, \aap, 562,
  A146

\bibitem[{{Jacobson} \& {Frebel}(2014)}]{Jacobson14}
{Jacobson}, H.~R., \& {Frebel}, A. 2014, Journal of Physics G Nuclear Physics,
  41, 044001

\bibitem[{{Ji} {et~al.}(2014){Ji}, {Frebel}, \& {Bromm}}]{Ji14}
{Ji}, A.~P., {Frebel}, A., \& {Bromm}, V. 2014, \apj, 782, 95

\bibitem[{{Ji} {et~al.}(2015){Ji}, {Frebel}, \& {Bromm}}]{Ji15}
---. 2015, \mnras, 454, 659

\bibitem[{{Jiang} \& {van den Bosch}(2015)}]{Jiang15}
{Jiang}, F., \& {van den Bosch}, F.~C. 2015, \mnras, 453, 3575

\bibitem[{{Jordi} {et~al.}(2006){Jordi}, {Grebel}, \& {Ammon}}]{Jordi06}
{Jordi}, K., {Grebel}, E.~K., \& {Ammon}, K. 2006, \aap, 460, 339

\bibitem[{{Kelson}(2003)}]{Kelson03}
{Kelson}, D.~D. 2003, \pasp, 115, 688

\bibitem[{{Kirby} {et~al.}(2013){Kirby}, {Cohen}, {Guhathakurta}, {Cheng},
  {Bullock}, \& {Gallazzi}}]{Kirby13b}
{Kirby}, E.~N., {Cohen}, J.~G., {Guhathakurta}, P., {et~al.} 2013, \apj, 779,
  102

\bibitem[{{Kirby} {et~al.}(2008){Kirby}, {Simon}, {Geha}, {Guhathakurta}, \&
  {Frebel}}]{Kirby08}
{Kirby}, E.~N., {Simon}, J.~D., {Geha}, M., {Guhathakurta}, P., \& {Frebel}, A.
  2008, \apjl, 685, L43

\bibitem[{{Koch} {et~al.}(2013){Koch}, {Feltzing}, {Ad{\'e}n}, \&
  {Matteucci}}]{Koch13}
{Koch}, A., {Feltzing}, S., {Ad{\'e}n}, D., \& {Matteucci}, F. 2013, \aap, 554,
  A5

\bibitem[{{Koch} {et~al.}(2014){Koch}, {Hansen}, {Feltzing}, \&
  {Wilkinson}}]{Koch14a}
{Koch}, A., {Hansen}, T., {Feltzing}, S., \& {Wilkinson}, M.~I. 2014, \apj,
  780, 91

\bibitem[{{Koch} {et~al.}(2008){Koch}, {McWilliam}, {Grebel}, {Zucker}, \&
  {Belokurov}}]{Koch08}
{Koch}, A., {McWilliam}, A., {Grebel}, E.~K., {Zucker}, D.~B., \& {Belokurov},
  V. 2008, \apjl, 688, L13

\bibitem[{{Koch} \& {Rich}(2014)}]{Koch14b}
{Koch}, A., \& {Rich}, R.~M. 2014, \apj, 794, 89

\bibitem[{{Koch} {et~al.}(2009){Koch}, {Wilkinson}, {Kleyna}, {Irwin},
  {Zucker}, {Belokurov}, {Gilmore}, {Fellhauer}, \& {Evans}}]{Koch09}
{Koch}, A., {Wilkinson}, M.~I., {Kleyna}, J.~T., {et~al.} 2009, \apj, 690, 453

\bibitem[{{Korobkin} {et~al.}(2012){Korobkin}, {Rosswog}, {Arcones}, \&
  {Winteler}}]{Korobkin12}
{Korobkin}, O., {Rosswog}, S., {Arcones}, A., \& {Winteler}, C. 2012, \mnras,
  426, 1940

\bibitem[{{Lattimer} \& {Schramm}(1976)}]{Lattimer76}
{Lattimer}, J.~M., \& {Schramm}, D.~N. 1976, \apj, 210, 549

\bibitem[{{Lee} {et~al.}(2013){Lee}, {Johnston}, {Tumlinson}, {Sen}, \&
  {Simon}}]{Lee13}
{Lee}, D.~M., {Johnston}, K.~V., {Tumlinson}, J., {Sen}, B., \& {Simon}, J.~D.
  2013, \apj, 774, 103

\bibitem[{{Lind} {et~al.}(2011){Lind}, {Asplund}, {Barklem}, \&
  {Belyaev}}]{Lind11}
{Lind}, K., {Asplund}, M., {Barklem}, P.~S., \& {Belyaev}, A.~K. 2011, \aap,
  528, A103

\bibitem[{{Lugaro} {et~al.}(2012){Lugaro}, {Karakas}, {Stancliffe}, \&
  {Rijs}}]{Lugaro12}
{Lugaro}, M., {Karakas}, A.~I., {Stancliffe}, R.~J., \& {Rijs}, C. 2012, \apj,
  747, 2

\bibitem[{{Martin} {et~al.}(2008){Martin}, {de Jong}, \& {Rix}}]{Martin08}
{Martin}, N.~F., {de Jong}, J.~T.~A., \& {Rix}, H.-W. 2008, \apj, 684, 1075

\bibitem[{{McConnachie}(2012)}]{McConnachie12}
{McConnachie}, A.~W. 2012, \aj, 144, 4

\bibitem[{{Norris} {et~al.}(2010{\natexlab{a}}){Norris}, {Wyse}, {Gilmore},
  {Yong}, {Frebel}, {Wilkinson}, {Belokurov}, \& {Zucker}}]{Norris10a}
{Norris}, J.~E., {Wyse}, R.~F.~G., {Gilmore}, G., {et~al.} 2010{\natexlab{a}},
  \apj, 723, 1632

\bibitem[{{Norris} {et~al.}(2010{\natexlab{b}}){Norris}, {Yong}, {Gilmore}, \&
  {Wyse}}]{Norris10b}
{Norris}, J.~E., {Yong}, D., {Gilmore}, G., \& {Wyse}, R.~F.~G.
  2010{\natexlab{b}}, \apj, 711, 350

\bibitem[{{Norris} {et~al.}(2013){Norris}, {Yong}, {Bessell}, {Christlieb},
  {Asplund}, {Gilmore}, {Wyse}, {Beers}, {Barklem}, {Frebel}, \&
  {Ryan}}]{Norris13}
{Norris}, J.~E., {Yong}, D., {Bessell}, M.~S., {et~al.} 2013, \apj, 762, 28

\bibitem[{{Placco} {et~al.}(2014){Placco}, {Frebel}, {Beers}, \&
  {Stancliffe}}]{Placco14}
{Placco}, V.~M., {Frebel}, A., {Beers}, T.~C., \& {Stancliffe}, R.~J. 2014,
  \apj, 797, 21

\bibitem[{{Roederer}(2013)}]{Roederer13}
{Roederer}, I.~U. 2013, \aj, 145, 26

\bibitem[{{Roederer} \& {Kirby}(2014)}]{Roederer14a}
{Roederer}, I.~U., \& {Kirby}, E.~N. 2014, \mnras, 440, 2665

\bibitem[{{Roederer} {et~al.}(2014){Roederer}, {Preston}, {Thompson},
  {Shectman}, \& {Sneden}}]{Roederer14b}
{Roederer}, I.~U., {Preston}, G.~W., {Thompson}, I.~B., {Shectman}, S.~A., \&
  {Sneden}, C. 2014, \apj, 784, 158

\bibitem[{{Roederer} {et~al.}(2010){Roederer}, {Sneden}, {Thompson}, {Preston},
  \& {Shectman}}]{Roederer10}
{Roederer}, I.~U., {Sneden}, C., {Thompson}, I.~B., {Preston}, G.~W., \&
  {Shectman}, S.~A. 2010, \apj, 711, 573

\bibitem[{{Safranek-Shrader} {et~al.}(2014){Safranek-Shrader},
  {Milosavljevi{\'c}}, \& {Bromm}}]{SafShrad14a}
{Safranek-Shrader}, C., {Milosavljevi{\'c}}, M., \& {Bromm}, V. 2014, \mnras,
  438, 1669

\bibitem[{{Salvadori} {et~al.}(2015){Salvadori}, {Sk{\'u}lad{\'o}ttir}, \&
  {Tolstoy}}]{Salvadori15}
{Salvadori}, S., {Sk{\'u}lad{\'o}ttir}, {\'A}., \& {Tolstoy}, E. 2015, \mnras,
  454, 1320

\bibitem[{{Schlafly} \& {Finkbeiner}(2011)}]{Schlafly11}
{Schlafly}, E.~F., \& {Finkbeiner}, D.~P. 2011, \apj, 737, 103

\bibitem[{{Shen} {et~al.}(2015){Shen}, {Cooke}, {Ramirez-Ruiz}, {Madau},
  {Mayer}, \& {Guedes}}]{Shen15}
{Shen}, S., {Cooke}, R.~J., {Ramirez-Ruiz}, E., {et~al.} 2015, \apj, 807, 115

\bibitem[{{Simon} {et~al.}(2010){Simon}, {Frebel}, {McWilliam}, {Kirby}, \&
  {Thompson}}]{Simon10}
{Simon}, J.~D., {Frebel}, A., {McWilliam}, A., {Kirby}, E.~N., \& {Thompson},
  I.~B. 2010, \apj, 716, 446

\bibitem[{{Simon} \& {Geha}(2007)}]{Simon07}
{Simon}, J.~D., \& {Geha}, M. 2007, \apj, 670, 313

\bibitem[{{Simon} {et~al.}(2011){Simon}, {Geha}, {Minor}, {Martinez}, {Kirby},
  {Bullock}, {Kaplinghat}, {Strigari}, {Willman}, {Choi}, {Tollerud}, \&
  {Wolf}}]{Simon11}
{Simon}, J.~D., {Geha}, M., {Minor}, Q.~E., {et~al.} 2011, \apj, 733, 46

\bibitem[{{Sneden}(1973)}]{Sneden73}
{Sneden}, C.~A. 1973, PhD thesis, The University of Texas at Austin.

\bibitem[{{Sobeck} {et~al.}(2011){Sobeck}, {Kraft}, {Sneden}, {Preston},
  {Cowan}, {Smith}, {Thompson}, {Shectman}, \& {Burley}}]{Sobeck11}
{Sobeck}, J.~S., {Kraft}, R.~P., {Sneden}, C., {et~al.} 2011, \aj, 141, 175

\bibitem[{{Strigari} {et~al.}(2008){Strigari}, {Bullock}, {Kaplinghat},
  {Simon}, {Geha}, {Willman}, \& {Walker}}]{Strigari08}
{Strigari}, L.~E., {Bullock}, J.~S., {Kaplinghat}, M., {et~al.} 2008, \nat,
  454, 1096

\bibitem[{{Vargas} {et~al.}(2013){Vargas}, {Geha}, {Kirby}, \&
  {Simon}}]{Vargas13}
{Vargas}, L.~C., {Geha}, M., {Kirby}, E.~N., \& {Simon}, J.~D. 2013, \apj, 767,
  134

\bibitem[{{Walsh} {et~al.}(2007){Walsh}, {Jerjen}, \& {Willman}}]{Walsh07}
{Walsh}, S.~M., {Jerjen}, H., \& {Willman}, B. 2007, \apjl, 662, L83

\bibitem[{{Walsh} {et~al.}(2008){Walsh}, {Willman}, {Sand}, {Harris}, {Seth},
  {Zaritsky}, \& {Jerjen}}]{Walsh08}
{Walsh}, S.~M., {Willman}, B., {Sand}, D., {et~al.} 2008, \apj, 688, 245

\bibitem[{{Wanajo} {et~al.}(2003){Wanajo}, {Tamamura}, {Itoh}, {Nomoto},
  {Ishimaru}, {Beers}, \& {Nozawa}}]{Wanajo03}
{Wanajo}, S., {Tamamura}, M., {Itoh}, N., {et~al.} 2003, \apj, 593, 968

\bibitem[{{Weisz} {et~al.}(2014){Weisz}, {Johnson}, \& {Conroy}}]{Weisz14c}
{Weisz}, D.~R., {Johnson}, B.~D., \& {Conroy}, C. 2014, \apjl, 794, L3

\bibitem[{{Wise} {et~al.}(2014){Wise}, {Demchenko}, {Halicek}, {Norman},
  {Turk}, {Abel}, \& {Smith}}]{Wise14}
{Wise}, J.~H., {Demchenko}, V.~G., {Halicek}, M.~T., {et~al.} 2014, \mnras,
  442, 2560

\bibitem[{{Yong} {et~al.}(2013){Yong}, {Norris}, {Bessell}, {Christlieb},
  {Asplund}, {Beers}, {Barklem}, {Frebel}, \& {Ryan}}]{Yong13a}
{Yong}, D., {Norris}, J.~E., {Bessell}, M.~S., {et~al.} 2013, \apj, 762, 26

\end{thebibliography}
